\documentclass[12pt]{article}
\usepackage{rotating}
\usepackage{epsfig}
\usepackage{psfig}
\usepackage[cp1251]{inputenc}
\usepackage{amsmath,bbm}
\usepackage{color}
\begin{document}
\newcommand{\be}{\begin{equation}}
\newcommand{\ee}{\end{equation}}

\newcommand{\cc}{\cite}
\newcommand{\ba}{\begin{eqnarray}}
\newcommand{\ea}{\end{eqnarray}}

%\end{document}
\begin{titlepage}
\begin{center}

\vspace{5cm}

  {\Large \bf Anomalous quark-gluon chromomagnetic interaction and high energy $\rho$-meson electroproduction  }
  \vspace{0.50cm}\\
 Nikolai Korchagin$^{a,}$\footnote{kolya.korchagin@gmail.com},  Nikolai Kochelev$^{a,}$\footnote{kochelev@theor.jinr.ru},
 Nikolai Nikolaev
$^{b,c,}$\footnote{n.nikolaev@fz-juelich.de} \vspace{0.50cm}\\
{(a) \it Bogoliubov Laboratory of Theoretical Physics,
Joint Institute for Nuclear Research, Dubna, Moscow region,
141980, Russia}\\
{(b)\it  L.D. Landau Institute for Theoretical  Physics,st. Kosygina 2, 119334 Moscow, Russia}\\
\vskip 1ex {(c) \it Forschungszentrum, Institute f\"ur Kernphysik, Postfach 1913, D-52425 J\"ulich, Germany } \vskip 1ex
\end{center}
\vskip 0.5cm \centerline{\bf Abstract}

It is shown that existence of a large anomalous chromomagnetic
moment of quark  induced by non-perturbative structure of QCD
leads to the additional contribution to exclusive $\rho$-meson
electroproduction off proton target. The significant contribution
coming from new type of quark-gluon interaction  to the $\rho
$-meson production cross section for both transversal and
longitudinal polarization of virtual photon is found. Such
non-perturbative contribution together with conventional
perturbative two-gluon exchange  allows us to describe the
experimental data at low $Q^2$ for transversal polarization.
However, in the longitudinal polarization case there is still some
discrepancy with the data. The possible source of this deviation
is discussed.

\vskip 0.3cm \leftline{Pacs: 24.85.+p, 12.38.-t, 12.38.Mh, 12.39.Mk}
 \leftline{Keywords: quarks, gluons, non-perturbative QCD, vector meson,
electroproduction}
\vspace{1cm}
\end{titlepage}
\setcounter{footnote}{0}

\section{Introduction}

One of salient features of perturbative high-energy QCD is
conservation of the $s$-channel helicity of quarks. On the other
hand, the QCD vacuum possesses a  nontrivial topological structure
- instantons are an extensively studied example (for the reviews
\cite{shuryak,diakonov}). Such topological fluctuations generates
the celebrated multiquark t'Hooft interaction which is
responsible, for example, for the solution of $U_A(1)$ problem in
QCD \cite{thooft}. Additionally, instantons were shown to generate
a non-trivial spin-flip, i.e., s-channel helicity non-conserving,
{\it quark-gluon interaction}
 \cite{kochelev1}. This interaction can be described in terms of
an anomalous chromomagnetic moment of quarks (ACMQ) complementary
to the perturbative Dirac coupling. Novel contributions from this
non-perturbative interaction to the soft Pomeron, gluon
distribution in the nucleon and sizable spin effects in strong
interactions have already been discussed in the literature
\cite{kochelev1,kochelev2,kochelev3,diakonov}. A magnitude of the
ACMQ can be related to the instanton density in the QCD vacuum.

The exclusive vector meson electroproduction is a unique testing
ground of the $s$-channel helicity properties of the quark-gluon
coupling. Specifically, $s$-channel helicity non-conserving
transitions from photons to vector mesons are possible even within
the perturbative QCD for a fundamental reason that a helicity of
mesons can be different from a sum of the quark and antiquark
helicities (see \cite{ivanovnikolaevsavin} and references
therein). The ACMQ would introduce an extra contribution to both
the $s$-channel helicity conserving and non-conserving vector
meson production amplitudes.
  Although it comes from manifestly soft
region, a direct evaluation of such a contribution is needed.

Helicity properties of electroproduced vector mesons have been
extensively studied in three experiments at HERA DESY, i.e. by the
H1, ZEUS and HERMES collaborations. Although gross features of
these data are well consistent with pQCD-based theoretical
predictions (see review  \cite{ivanovnikolaevsavin} and recent
development in \cite{pire1},\cite{pire2} ), there remain several
open issues. An outstanding problem is a large relative  phase of
the leading helicity amplitudes for production of longitudinal and
transverse $\rho^0$'s and $\phi$'s, which has been observed by the
HERMES and H1 Collaborations \cite{HERMES1,HERMES2,H1}. It is
definitely larger than the prediction based on the handbag
mechanism \cite{goloskokovkroll} and also in the conventional pQCD
pomeron-based  color dipole approach \cite{dipole} (see discussion
in \cite{ivanovnikolaevsavin}). Besides that, pQCD-driven
approaches seem to fail with the experimentally observed $Q^2$
dependence of the ratio $\sigma_L/\sigma_T$ in full experimentally
studied range of $Q^2$: theoretical calculations substantially
overestimate  this ratio at large $Q^2$. Finally, it is important
to evaluate an impact of new non-perturbative mechanism on the
transition from real to virtual electroproduction.

In the present  paper we   study exclusive $\rho$- meson
electroproduction off protons with allowance for the novel
$s$-channel helicity non-conservation mechanism  induced by the
anomalous chromomagnetic moment of quarks. We focus on the
simplest observables, $\sigma_L$ and $\sigma_T$, a calculation of
the full set of helicity amplitudes will be reported elsewhere.

\section{ Anomalous quark-qluon chromomagnetic interaction}

In the most general case, the interaction vertex
of massive quark with gluon can be written in the following form:
\begin{equation}
V_\mu(k_1^2,k_2^2,\kappa^2)t^a = -g_st^a[F_1(k_1^2,k_2^2,\kappa^2) \gamma_\mu
 +
\frac{\sigma_{\mu\nu}\kappa_\nu}{2m}F_2(k_1^2,k_2^2,\kappa^2)],
 \label{vertex}
 \end{equation}
where the first term is a conventional perturbative QCD
quark-gluon vertex and the second term comes from non-perturbative
sector of QCD. In Eq.\ref{vertex} the form factors $F_{1,2}$
describe a nonlocality of the non-perturbative interaction,
$k_{1,2}$ are the momenta
 of incoming and outgoing
quarks, respectively, and $ \kappa=k_2-k_1$, $m$ is the quark
mass, and $\sigma_{\mu\nu}=(\gamma_\mu \gamma_\nu-\gamma_\nu
\gamma_\mu)/2$.
 In what follows, we focuse on effects of the novel
color chromomagnetic vertex and keep $F_1(k_1^2,k_2^2,\kappa^2)=1$.
The anomalous quark chromomagnetic moment
(AQCM) equals
\begin{equation}
\mu_a=F_2(0,0,0). \nonumber
\end{equation}

In the earlier paper \cite{kochelev1} it was
shown that instantons, a strong vacuum fluctuations of gluon
fields with nontrivial topology,
generate an  ACMQ  which is proportional to the  instanton density
 \be
\mu_a=-\pi^3\int \frac{d\rho n(\rho)\rho^4}{\alpha_s(\rho)}.
\nonumber\ee
In terms of the average size of instantons $\rho_c$ and the
 dynamical quark mass $m$ in
non-perturbative QCD vacuum one finds \cite{kochelev2}
 \be
\mu_a=-\frac{3\pi (m\rho_c)^2}{4\alpha_s(\rho_c)},
 \label{mu} \ee
which exhibits a strong sensitivity of the ACMQ to a dynamical
mass of quarks. To this end we emphasize an implicit assumptions
that quarks are light, ie., the above estimates of ACQM are valid
for $u,d,s$, while for heavy quarks ACQM vanishes. For the average
instanton size $\rho_c^{-1}=0.6$ GeV
   this mass is  changing from
  $m=170 MeV$
in the mean field (MF) approximation  to
  $m=345$ MeV   within Diakonov-Petrov (DP) model.
The QCD strong coupling constant at the instanton scale  can be
evaluated as \be \alpha_s(\rho_c)\approx 0.5, \nonumber \ee  and
the resulting AQCM for light quarks is numerically quite large:
\be \mu_a^{MF}\approx-0.4, \ \ \
\mu_a^{DP}\approx-1.6.\nonumber\ee Recently,  an AQCM of similar
magnitude has been obtained within the Dyson-Schwinger equation
approach to non-perturbative QCD (see discussion and references
in\cite{Chang:2011vu}).

The form factor  $F_2(k_1^2,k_2^2,\kappa^2)$
 suppresses the AQCM vertex
at short distances when the respective virtualites are large. Within the
instanton model its explicit is related to Fourier-transformed quark
zero-mode and instanton fields and reads
\be
 F_2(k_1^2,k_2^2,\kappa^2) =\mu_a
\Phi_q(\mid k_1\mid\rho/2)\Phi_q(\mid k_2\mid\rho/2)F_g(\mid
\kappa\mid\rho) \ , \nonumber \ee where \ba
\Phi_q(z)&=&-z\frac{d}{dz}(I_0(z)K_0(z)-I_1(z)K_1(z)), \nonumber\\
F_g(z)&=&\frac{4}{z^2}-2K_2(z) \nonumber \ea
 are the
where  $I_{\nu}(z)$, $K_{\nu}(z)$, are the modified
Bessel functions and  $\rho$  is the instanton size.

Recent discussion has shown \cite{kochelev2} that the ACQM
contribution complements the pQCD evaluations of the  total
quark-quark cross section in a way which improves constituent
quark model description of high energy nucleon-nucleon total cross
section. Furthermore, such model provide the soft contribution to
the gluon distribution in the nucleon which is consistent with
initial conditions to a DGLAP evolution of phenomenological PDFs.
A purpose of the present communication is to explore the effects
of ACQM in elastic (diffractive) electroproduction of $\rho$
mesons.

\section{Model for vector meson exclusive production}

Driven by VBKL considerations, high energy diffractive production of
vector mesons is usually desctribed by an exchange of a color-singlet
two-gluon tower in the $t$-channel. The corresponding diagrams
are presented in Fig.1.
\begin{figure}[h]
%\hspace*{5cm}
\centerline{\epsfig{file=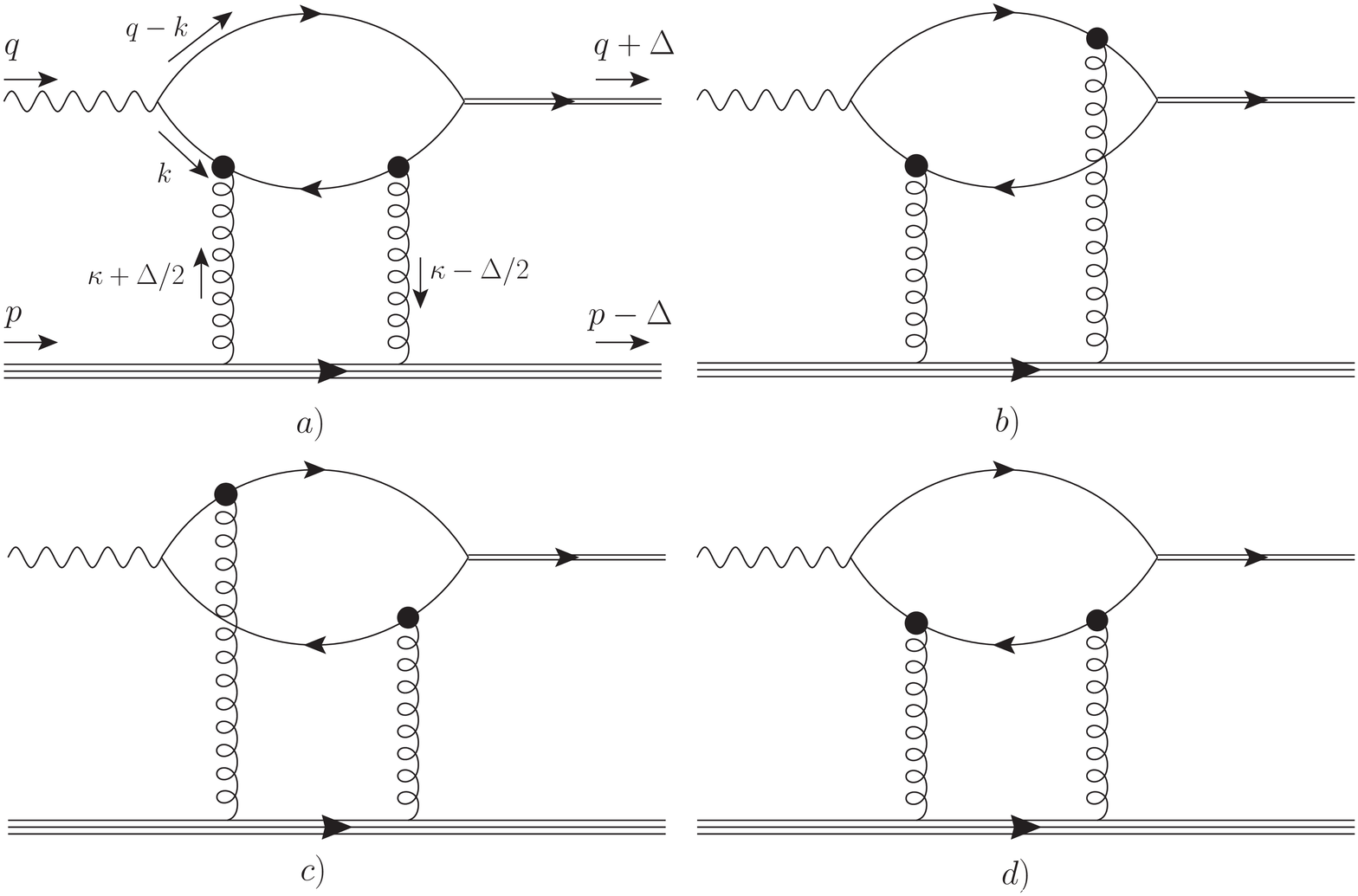,width=14cm,height=10cm}}
%\hspace*{5cm}
\caption{The diagrams which contribute to high energy exclusive vector meson electroproduction off proton by
two gluon  exchange. Here the blob stands for the
generalized quark-gluon  vertex Eq.\ref{vertex}.}
\label{4diagrams}
\end{figure}
By using the Sudakov expansion for the momenta of proton and virtual photon
\begin{eqnarray}
&&p_{\mu}=p'_{\mu}+\frac{m^2_p}{s}q'_{\mu}, \quad   q_{\mu}=q'_{\mu}-x p'_{\mu}, \quad q'^2=p'^2=0,
\end{eqnarray}
where $Q^2=-q^2$,  $x=\frac{Q^2}{s}\ll 1$, $s=2(p'q')$,  the quark momentum  $k$  in the quark loop, gluon momentum $\kappa$, and momentum transfer to proton $\Delta$ (see Fig.1) can be presented in the following form
\begin{eqnarray}
   k_{\mu}&=&yp^\prime_{\mu}+zq^\prime_{\mu}+\vec{k}_\mu, \notag \\
    \kappa_{\mu}&=&\alpha p^\prime_{\mu}+\beta q^\prime_{\mu}+\vec{\kappa}_{\mu}, \\
    \Delta_{\mu}&=&\delta p^\prime_{\mu}+\sigma q^\prime_{\mu}+\vec{\Delta}_{\mu},\notag
\end{eqnarray}
where any vector $\vec{l}$  is transversal part of four-vector $l_\mu$ which  satisfies   the relation
 $\vec{l}\cdot q^\prime=\vec{l}\cdot p^\prime=0$.

Hereafter we follow the $k$-factorization analyis devloped in
\cite{Kuraev:1998ht,Ivanov:2003iy,ivanovnikolaev}.
The polarization vectors for virtual photon $e$ and vector meson $V$ are
\begin{eqnarray}
&&e_{T \mu}=\vec{e}_{\mu}, \notag\\
&&e_{L \mu}=\frac{1}{Q}(q^\prime_{\mu}+x p^\prime_{\mu}),\notag\\
&&V_{T\mu}=\vec{V}_{\mu}+\frac{2(\vec{\Delta }\vec{V})}{s}p^\prime_{\mu},\\
&&V_{L \mu}=\frac{1}{M}\left(q^\prime_{\mu}+\frac{\vec{\Delta}^2-M^2}{s} p^\prime_{\mu}+\vec{\Delta}_{\mu} \right)\notag
\end{eqnarray}
where $M$ is the mass of $\bar q q$ pair on mass-shell:
\begin{equation}
    M^2=\frac{\vec{k}^2+m^2}{z(1-z)}
\end{equation}
The imaginary part of the amplitude takes the form
\begin{eqnarray}
A(x,Q^{2},\vec \Delta)&=&
- is{c_{V}\sqrt{4\pi\alpha_{em}}
\over 4\pi^{2}}
\int {dz\over z(1-z)} \int d^2 \vec k \psi(z,\vec k) \int {d^{2} \vec \kappa
\over
\vec\kappa^{4}}\alpha_{S}{\cal{F}}(x,\vec \kappa,\vec \Delta) \times \nonumber
\\
&&\times \biggl[ \frac{1-z}{z}\frac{I^{(a)}}{\vec{k}_{1a}^2+m^2+z(1-z)Q^2}+
\frac{I^{(b)}}{\vec{k}_{1b}^2+m^2+z(1-z)Q^2}+  \\
&&\quad +\frac{I^{(c)}}{\vec{k}_{1c}^2+m^2+z(1-z)Q^2} +
\frac{z}{1-z}\frac{I^{(d)}}{\vec{k}_{1d}^2+m^2+z(1-z)Q^2}\biggr] \nonumber.
\end{eqnarray}

Here $\alpha_{em}$ is the fine-structure constant,
$c_V=1/\sqrt{2}$ is coming from the flavor part of the $\rho$
meson wave function, ${\cal{F}}(x,\vec \kappa,\vec \Delta)$ is the
differential gluon density and $\psi(z,\vec k)$ is light-cone wave
function of the $\rho$-meson, for which we us a simple
parameterization \begin{equation}
    \psi=c \exp \left(-\frac{a^2 \mathbf{p}^2}{2}\right) =c \exp \left( -\frac{a^2}{2} \left( \vec{k}^2+\frac{1}{4}(2z-1)^2M^2\right) \right)
\end{equation}
where $\mathbf{p}$ is 3-dimensional relative momentum of quarks in  pair. Two constants $a$ and $c$ were fixed  by the
normalization of wave function and  decay width $\Gamma(\rho \to e^+ e^-)$, we
find $a=3.927 \textrm{ GeV}^{-1}$, $c=17.44$.

For the QCD running coupling we use
\begin{equation}
\alpha_s(q^2)=\frac{4\pi}{9\ln((q^2+m_g^2)/\Lambda_{QCD}^2)},
\end{equation}
where $\Lambda_{QCD}=0.28$ GeV  and the value $m_g=0.88$ GeV
imposes the infrared freezing at $\alpha_s(1/\rho_c^2)\approx
\pi/6$ \cite{diakonov}. Here $q^2$ is the maximum virtuality of
momentum which inserted to vertex, i.e.
$q^2=\textrm{Max}(k_1^2,k_2^2,\kappa^2)$.

$I^{(i)}$ is the trace over quark line in $i)$ diagram from Fig.\ref{4diagrams} divided by $2s^2$. It is involved a three parts:
\begin{equation}
    I^{(i)}=I_{pert}^{(i)}+I_{cm}^{(i)}+I_{mix}^{(i)}.
\end{equation}

Below we are using the following notation for the transverse momentum of quark
\begin{eqnarray}
    \vec{k}_{1a}=\vec{k}-(1-z)\vec{\Delta} \notag \\
    \vec{k}_{1b}=\vec{k}-(1-z)\vec{\Delta}+\vec{\kappa}+\frac{1}{2} \vec{\Delta} \notag \\
    \vec{k}_{1c}=\vec{k}-(1-z)\vec{\Delta}-\vec{\kappa}+\frac{1}{2} \vec{\Delta}  \\
    \vec{k}_{1d}=\vec{k}+z\vec{\Delta} \notag
\end{eqnarray}
Such shift of momentum is needed for keeping transverse momentum of quarks inserted to meson vertex to be equal $\vec k$ for all diagrams in order to take out $\psi(z,\vec k)$ as common multiplier. Besides, $[\vec a \vec b]=a_x b_y-a_y b_x$.

The formula for $TT$ transition in case when both of quark-gluon vertices are perturbative is
\begin{eqnarray}
    I_{pert}^{(c)}(T \rightarrow T) &=&\left[(\vec{e}\vec{V}^*)(m^2+\vec{k}\vec{k}_{1c})+(\vec{V}^*\vec{k}) (\vec{e}\vec{k}_{1c})(1-2z)^2 -(\vec{e}\vec{k})(\vec{V}^*\vec{k}_{1c})\right] .
\end{eqnarray}

For pure perturbative vertices there is the relation between different contributions
  \begin{equation}
  I_{pert}^{(b)}=I_{pert}^{(c)}=-\frac{1-z}{z}I_{pert}^{(a)}=-\frac{z}{1-z}I_{pert}^{(d)}
  \nonumber
  \end{equation}
subject to a proper substitution of the loop quark momenta the relevant
diagrams of Fig.1.

When both quark-gluon vertices come from nonperturbative ACQM, we find
\begin{eqnarray}
    I_{cm}^{(a)}(T \rightarrow T) &=&-\frac{z}{1-z}\left[(\vec{e}\vec{V}^*)(m^2+\vec{k}\vec{k}_{1a})+(\vec{V}^*\vec{k}) (\vec{e}\vec{k}_{1a})(1-2z)^2 -(\vec{e}\vec{k})(\vec{V}^*\vec{k}_{1a})\right] \vec \kappa^2 \times \notag \\ && \times F_2(k_{I\,\text{avg}}^2,0,\kappa^2)F_2(k_{I\!I\,\text{avg}}^2,0,\kappa^2).
\end{eqnarray}
Here we notice that one of quarks is always on mass-shell, and
$k_{I,II\,\text{avg}}^2$ stand for the virtuality of the off-mass shell quark.
To obtain formula for $I_{cm}^{(d)}$ one should perform the substitutions  $z/(1-z) \to (1-z)/z $ and $\vec k_{1a} \to \vec k_{1d}$ in above expression.
The AQCM contribution from graph in Fig.1c is
\begin{eqnarray}
 I_{cm}^{(c)}(T \rightarrow T)=\Big[ \big( (1-2z)^2(\vec{V}^*\vec{k})(\vec{k}_{1c}\vec{e}) -(\vec{k}\vec{k}_{1c})(\vec{e}\vec{V}^*)+ (\vec{k}\vec{e})(\vec{V}^*\vec{k}_{1c})\big)\vec{\kappa}^2+ \notag \\ +m^2\big(2(\vec e \vec \kappa)(\vec V^* \vec \kappa)-(\vec e \vec V^*)\vec{\kappa}^2 \big) \Big] F_2(k_{I\,\text{avg}}^2,0,\kappa^2)F_2(k_{I\!I\,\text{avg}}^2,0,\kappa^2).
\end{eqnarray}
The replacement $\vec k_{1c} \to \vec k_{1b}$ leads to the formula for  $I_{cm}^{(b)}$.

The interference of the pQCD and ACQM vertices gives
\begin{eqnarray}
   I_{mix}^{(c)}(T \rightarrow T)=m
   \left[(\vec{V}^*\vec{k})(\vec{\kappa}\vec{e})(1-2z) - [\vec{e}\vec{\kappa}]
     [ \vec{V}^*\vec{k}] - (\vec{e}\vec{k}_{1c})(\vec{\kappa}\vec{V})(1-2z) -
     [\vec{e}\vec{k}_{1c}] [ \vec{\kappa}\vec{V}^*]  \right] \notag \times \\
   \times
   (F_2(k_{I\!I\,\text{avg}}^2,0,\kappa^2)-F_2(k_{I\,\text{avg}}^2,0,\kappa^2)) \label{formula}
\end{eqnarray}
The $I_{mix}^{(b)}$ is obtained from the above by substitution  $k_{1c} \to
k_{1b}$. The remaining amplitudes are

\begin{eqnarray}
   I_{mix}^{(a)}(T \rightarrow T)&=&\frac{z m}{1-z} \left[(\vec{V}^*\vec{k})(\vec{\kappa}\vec{e})(1-2z) - [\vec{e}\vec{\kappa}] [ \vec{V}^*\vec{k}] - (\vec{e}\vec{k}_{1a})(\vec{\kappa}\vec{V})(1-2z) - [\vec{e}\vec{k}_{1a}] [ \vec{\kappa}\vec{V}^*]  \right] \times \notag \\ && \times \left(F_2(k_{I\!I\,\text{avg}}^2,0,\kappa^2) -F_2(k_{I\,\text{avg}}^2,0,\kappa^2) \right)
\end{eqnarray}

\begin{eqnarray}
   I_{mix}^{(d)}(T \rightarrow T)&=&\frac{(1-z) m}{z} \left[(\vec{V}^*\vec{k})(\vec{\kappa}\vec{e})(1-2z) + [\vec{e}\vec{\kappa}] [ \vec{V}^*\vec{k}] - (\vec{e}\vec{k}_{1d})(\vec{\kappa}\vec{V}^*)(1-2z) + [\vec{e}\vec{k}_{1d}] [ \vec{\kappa}\vec{V}^*]  \right] \times \notag \\ && \times \left(F_2(k_{I\!I\,\text{avg}}^2,0,\kappa^2) -F_2(k_{I\,\text{avg}}^2,0,\kappa^2) \right).
\end{eqnarray}

For the longitudinal polarization we obtain
\begin{eqnarray}
I_{pert}^{(c)}(L \rightarrow L) = -4 Q M z^2 (1-z)^2\notag \\
I_{cm}^{(c)}(L \rightarrow L) = -4 Q M z^2 (1-z)^2 \vec \kappa ^2 F_2(k_{I\,\text{avg}}^2,0,\kappa^2)F_2(k_{I\!I\,\text{avg}}^2,0,\kappa^2).
\end{eqnarray}
And $I^{(b)}=I^{(c)}=-\frac{1-z}{z}I^{(a)}=-\frac{z}{1-z}I^{(d)}$ with corresponding $\vec k_{1i}$.
In the longitudinal case the contribution from the interference of pQCD and
AQCM vertices
vanishes
\begin{equation}
I_{mix}^{(a)}(L \rightarrow L)=I_{mix}^{(b)}=I_{mix}^{(c)}=I_{mix}^{(d)} = 0.
\end{equation}
%\vskip 2cm

A common feature of nonperturbative approaches is a fast running dynamical
mass of constiituent quarks which drops to a small current quark mass at
large virtualities. In principle, the running of the quark masses affect the
$Q^2$ dependence of vector meson production observables. However, such an fill-fledged
 involved calculation is beyond the scope
 of
the present communication. Here we only recall that
according to the color transparency
considerations the vector meson production amplitudes are dominated by
the components of the vector meson wave function taken at transverse size,
i.e., the scanning radius $r_S \sim 6/\sqrt{Q^2 + m_V^2}$ \cite{nikolaevadd,kopeliovich,ivanovnikolaevsavin}.
 Consequently, the
virtuality of quarks is $\propto (Q^2 + m_V^2)$, and arguably
one can model an approach
to the pQCD regime at large virtuality of the  photon making use
of a simple approximation
\begin{equation}
    m(Q^2)=\frac{m(0)}{1+Q^2/m_V^2},
\label{run}
\end{equation}
where $m(0)=345\textrm{ MeV}$ and $m_V=770\textrm{ MeV}$

\section{Discussion of the results for $\sigma_L$ and $\sigma_T$}

The final result for cross sections is presented in Fig.2 in comparison with data obtained
by H1 and ZEUS Collaborations.
Helicity flip transitions are found to give a
 very small contribution to the total production
cross section, see also discussion in \cite{ivanovnikolaevsavin}:
\begin{eqnarray}
    \sigma_T=\sigma_{T\to T}+\sigma_{T\to L} \approx \sigma_{T\to T} \\
\sigma_L=\sigma_{L\to L}+\sigma_{L\to T} \approx \sigma_{L\to L}. \notag
\end{eqnarray}
\vskip 6cm
\hspace*{-2cm}
\begin{figure}[htb]
\begin{minipage}[c]{9cm}
%\hskip 1cm
 \vskip -6.0cm
\includegraphics[scale=0.6]{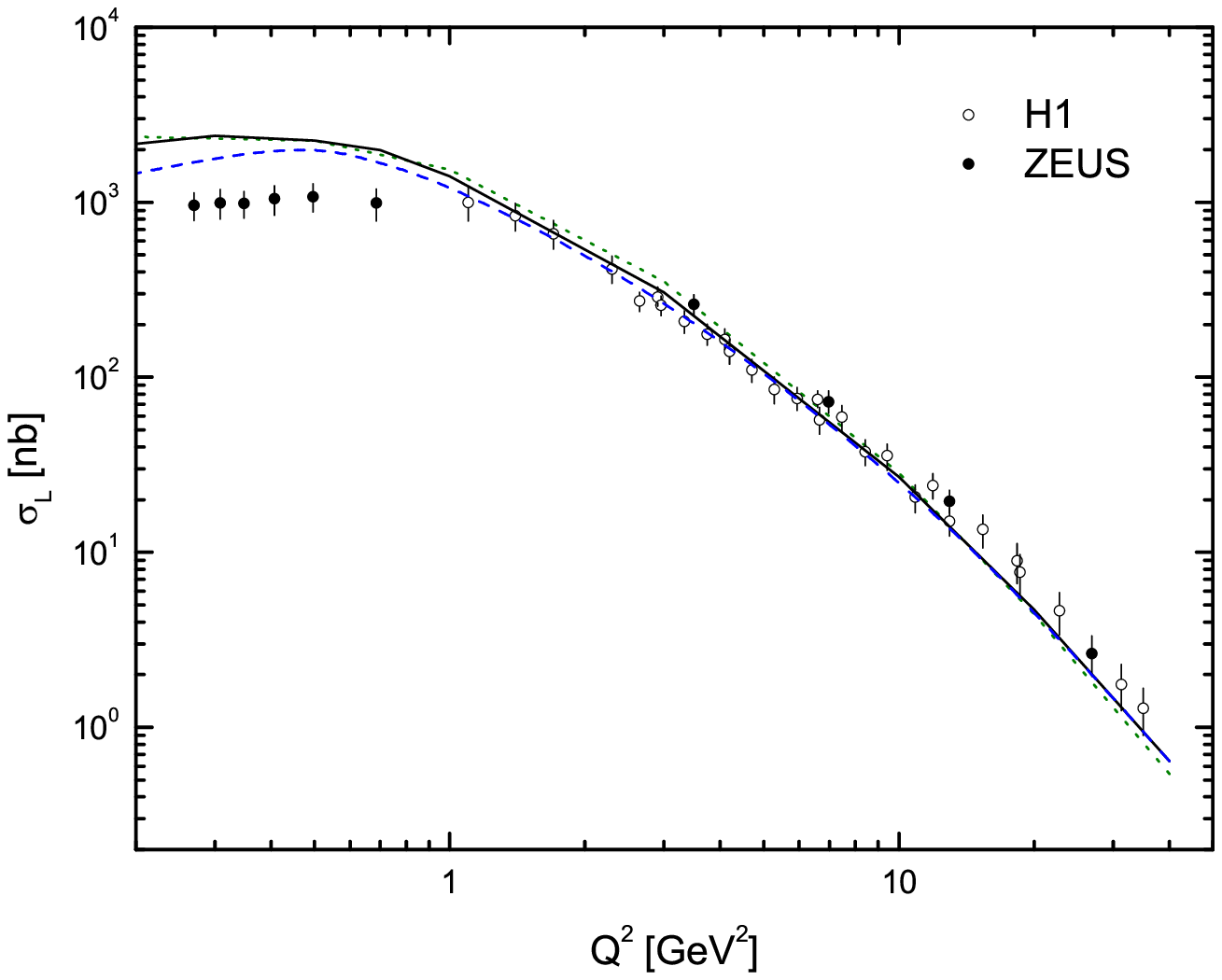}
%\centerline{\epsfig{file=longcor.eps,width=6cm,height=6cm}}
\end{minipage}
\begin{minipage}[c]{9cm}
%\centerline{\epsfig{file=csT.eps,width=8cm,height=6cm}}
\hspace*{-1cm}
 \vskip -6cm
\includegraphics[scale=0.6]{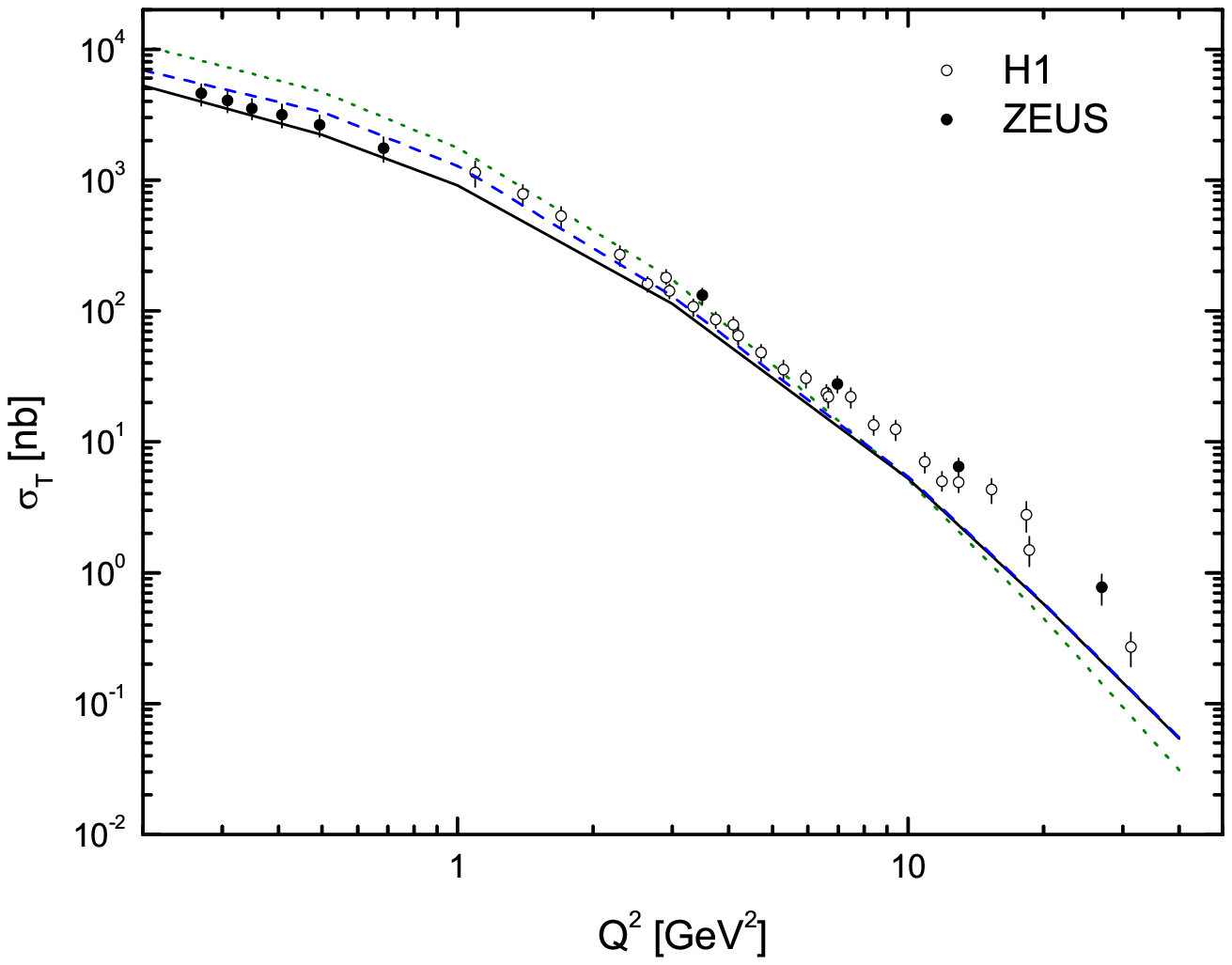}
%\centerline{\epsfig{file=transcor.eps,width=6cm,height=6cm}}
%\hskip -2cm
\end{minipage}
%\vskip -2cm
\caption{Cross sections of $\rho$ meson exclusive
electroproduction: left panel is for longitudinal virtual photon
polarization and right panel for transversal photon polarization.
The solid line is the calculation with AQCM, dashed line is the
result of  pQCD contribution with running quark mass,
Eq.\ref{run}, and dotted line is pQCD calculation with fixed quark
mass $m_q=220$ MeV  \cite{ivanovnikolaev,Ivanov:2003iy}.
Experimental points are taken from  \cite{Aaron:2009xp} for H1
and from \cite{Clerbaux:1999sn} for ZEUS Collaborations.}
\end{figure}

The longitudinal cross-section $\sigma_L$ is free of the
interference of the pQCD and AQCM vertices.  Allowance for AQCM
effects slightly enhances  $\sigma_L$ in the region of
non-perturbative small $Q^2$, but still the cross section is
overestimated in low $Q^2$ region.  There is one caveat, though:
we evaluated the pQCD contribution using the unintegrated gluon
density which has been tuned to the experimental data on the
proton structure function. To be more accurate, one must reanalyze
the structure function data with allowance for the AQCM effect,
arguably that would lower somewhat the resulting pQCD contribution
to $\sigma_L$ bringing the theoretical curve closer to the
experimental data points.

\begin{figure}[h]
%\vspace*{-3cm}
%\hspace*{3cm}
\centerline{\epsfig{file=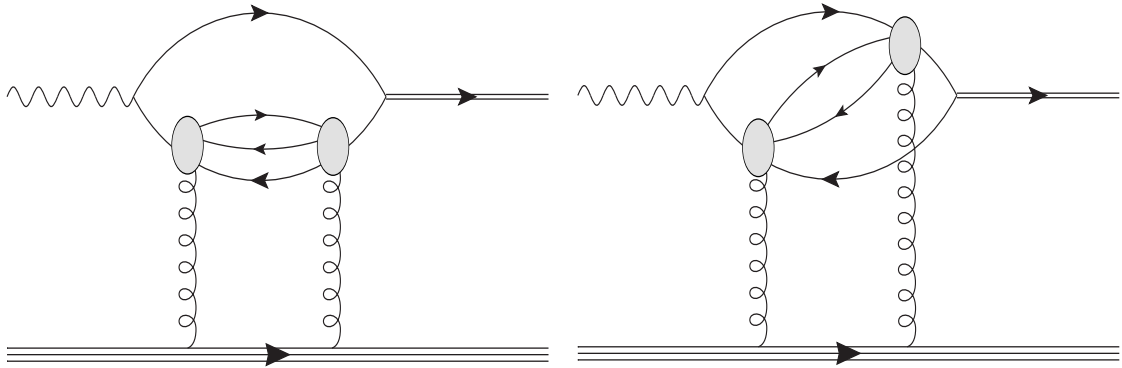,width=15cm,height=5cm}}
%\hspace*{5cm}
\caption{Additional contribution to exclusive electroproduction
 of the $\rho$ meson coming from quark-antiquark exchange between chromomagnetic
vertices.}
\end{figure}

The case of the transverse cross section $\sigma_T$ is more
subtle. In this case the effect of the pQCD-AQCM interference is
quite substantial. As a matter of fact, the resulting destructive
interference numerically takes over the pure AQCM contribution and
lowers $\sigma_T$ compared to the pure pQCD contribution. It is
well understood that $\sigma_T$ is more susceptible to the
non-perturbative effects in comparison with $\sigma_L$. Indeed,
the non-perturvative corrections to $\sigma_T$ die out
substantially slower than corrections to $\sigma_L$.

To this end we notice that we only treated a leading in $1/N_c$
restricted class of the instanton induced non-perturbative QCD
interactions which are reducible to the anomalous chromomagnetic
quark-glion vetrex. The full-fledged t'Hooft's-like multipartonic
interaction \cite{thooft} gives rise to a more complicated
diagrams,  examples are shown in Fig.3.  The issue of such
contributions will be considered elsewhere
\cite{korchaginkochelevnikolaev}. We also have checked the
possible effect of introduction of the form factor into pQCD
vertex which cuts low transfer momentum region where one-gluon
exchange picture looks questionable. It leads to the decreasing of
the both longitudinal and transverse cross sections  at low $Q^2$.
In this case we observe significant improvement of agreement  with
experiment of our calculation for $\sigma_L$, but agreement of
calculated $\sigma_T$ with data becomes worse.

As we  emphasized, the ACQM vertex manifestly violates the quark
$s$-channel helicity conservation. Arguably, the effects of ACQM
will be stronger in the helicity flip amplitudes of vector meson
electroproduction. The numerical results for full spin desnity
matrix of diffractive $\rho$-mesons will be presented elsewhere.

\section{Acknowledgments}
The authors are very grateful to I.~O.~Cherednikov, A.E. Dorokhov,
  E.A.~Kuraev and L.N. Lipatov
   for useful
discussions. This work was supported in part   by  RFBR grant 10-02-00368-a, by  Belarus-JINR grant,
and by Heisenberg-Landau program.

\end{document}